\documentclass[11pt,letterpaper]{article}

\usepackage{amsmath,amsthm,amssymb}
\usepackage{graphicx}
\usepackage{xcolor}
\usepackage{enumitem}
\usepackage{url}
\usepackage{epsfig}
\usepackage{hyperref}
\usepackage{geometry}
\usepackage[numbers]{natbib}

\newtheorem{theorem}{Theorem}
\newtheorem{lemma}{Lemma}
\newtheorem{example}{Example}
\newtheorem{fact}{Fact}

\newcommand{\INPUT}{{\bf input:}\ }
\newcommand{\OUTPUT}{{\bf output:}\ }
\newcommand{\WHILE}{{\bf while}\ }
\newcommand{\DO}{{\bf do}\ }
\newcommand{\ENDWHILE}{{\bf end while}\ }

\begin{document}

\title{Sponsored Search, Market Equilibria, and the Hungarian Method\thanks{A preliminary version of this paper appeared in \citet{DHW10}.}}

\author{%
	Paul D\"{u}tting\thanks{%
	Ecole Polytechnique F\'ed\'erale de Lausanne (EPFL), 
	Station 14, CH-1015 Lausanne, Switzerland, 
	Email: \texttt{paul.duetting@epfl.ch}.}
	\and 
	Monika Henzinger\thanks{%
	University of Vienna, Faculty of Computer Science, 
	W\"ahringer Stra{\ss}e 29/6.32, A-1090 Vienna, Austria,
	Email: \texttt{monika.henzinger@univie.ac.at}.} 
	\and 
	Ingmar Weber\thanks{%
	Yahoo! Research Barcelona, Avinguda Diagonal 177 (8th Floor), 
	E-08018 Barcelona, Spain,
	Email: \texttt{ingmar@yahoo-inc.com}.}
}

\date{}

\maketitle

\begin{abstract}
Matching markets play a prominent role in economic theory. A prime example of such a market is the sponsored search market. Here, as in other markets of that kind, market equilibria correspond to feasible, envy free, and bidder optimal outcomes. For settings without budgets such an outcome always exists and can be computed in polynomial-time by the so-called Hungarian Method. Moreover, every mechanism that computes such an outcome is incentive compatible. We show that the Hungarian Method can be modified so that it finds a feasible, envy free, and bidder optimal outcome for settings with budgets. We also show that in settings with budgets no mechanism that computes such an outcome can be incentive compatible for all inputs. For inputs in general position, however, the presented mechanism---as any other mechanism that computes such an outcome for settings with budgets---is incentive compatible.
\end{abstract}

\section{Introduction}\label{sec:intro}
In a matching market $n$ bidders have to be matched to $k$ items. A prime example of such a market is the sponsored search market, where bidders correspond to advertisers and items correspond to ad slots. In this market each bidder has a per-click valuation $v_i$, each item $j$ has a click-through rate $\alpha_j$, and bidder $i$'s valuation for item $j$ is $v_{i,j} = \alpha_j \cdot v_i.$ More generally, each bidder $i$ has a valuation $v_{i,j}$ for each item $j$. In addition, each item $j$ has a reserve price $r_j$. A mechanism is used to compute an outcome $(\mu,p)$ consisting of a matching $\mu$ and per-item prices $p_j$. The bidders have quasi-linear utilities. That is, bidder $i$'s utility is $u_i = 0$ if he is unmatched and it is $u_i = v_{i,j} - p_j$ if he is matched to item $j$ at price $p_j$. The valuations are private information and the bidders need not report their true valuations if it is not in their best interest to do so.

Ideally, the market should be in equilibrium. In the context of matching markets this typically means that the outcome computed by the mechanism should be {\em feasible}, {\em envy free}, and {\em bidder optimal}. An outcome is feasible if all bidders have non-negative utilities and if the price of all matched items is at least the reserve price. It is envy free if it is feasible and if at the current prices no bidder would get a higher utility if he was assigned a different item. It is bidder optimal if it is envy free and if the utility of every bidder is at least as high as in every other envy free outcome. Another requirement is that the mechanism should be incentive compatible. A mechanism is incentive compatible if each bidder maximizes his utility by reporting truthfully no matter what the other bidders report.

For matching markets of the above form a bidder optimal outcome always exists \cite{SS72}, can be computed in polynomial time by the so-called Hungarian Method \cite{K55}, and every mechanism that computes such an outcome is incentive compatible \cite{L83}. The above model, however, ignores the fact that in practice bidders often have budgets. Concrete examples include Google's and Yahoo's ad auction. Budgets are also challenging theoretically as they lead to discontinuous utility functions and thus break with the quasi-linearity of the original model without budgets.

In our model each bidder can specify a maximum price for each item. If bidder $i$ specifies a maximum price of $m_{i,j}$ for item $j$, then he cannot pay any price $p_j \ge m_{i,j}$. Hence the utility of bidder $i$ is $u_i = 0$ if he is unmatched, it is $u_i = v_{i,j} - p_j$ if he is matched to item $j$ at price $p_j < m_{i,j}$ (strict inequality), and it is $u_i = - \infty$ otherwise.%
\footnote{While requiring $p_j \le m_{i,j}$ seems to be more intuitive, it has the disadvantage that the infimum envy prices may not be envy free themselves: There are three bidders and one item. All bidders have a valuation of $10$ and the first two bidders have a maximum price of $5$. Then any price $p \le 5$ is not envy free because all bidders would prefer to be matched, and any price $p > 5$ is not bidder optimal because a slightly lower price would still be envy free.}
As before an outcome is feasible if all bidders have non-negative utilities and if the price of all matched items is at least the reserve price. It is envy free if it is feasible and if at the current prices no bidder would get a higher utility if he was assigned a different item. It is bidder optimal if it is envy free and if the utility of every bidder is at least as high as in every other envy free outcome.

For this model we show that the Hungarian Method can be modified so that it always finds a bidder optimal outcome in polynomial time. We also show that no mechanism that computes such an outcome is incentive compatible for all inputs. For inputs in general position, i.e., inputs with the property that in a certain weighted multi-graph defined on the basis of the input no two walks have exactly the same weight, our mechanism---as any other mechanism that computes a bidder optimal outcome---is incentive compatible \cite{DHW11}. All our results can be extended to more general (but still linear) utility functions.

A similar problem was previously considered by \cite{AMPP09}. Their model differs from our model in several ways: 
(1) The utility $u_i$ of bidder $i$ is $u_i = 0$ if he is unmatched, it is $u_i = v_{i,j} - p_j$ if he is matched to item $j$ at price $p_j \le m_{i,j}$ (weak inequality), and it is $u_i = -\infty$ otherwise. 
(2) The reserve prices $r_{i,j}$ may depend on the bidders and the items.
(3) An outcome is envy free if it is feasible and if for all bidders $i$ and all items $j$ either (a) $u_i \ge v_{i,j} - \max(p_j,r_{i,j})$ or (b) $p_j \ge m_{i,j}$.
For these definitions they showed that for inputs in general position (a) a bidder optimal outcome always exists, (b) a bidder optimal outcome can be computed by a (rather complicated) mechanism in polynomial time, and (c) this mechanism is incentive compatible. For inputs that are not in general position a bidder optimal outcome may not exist as the following example shows.\footnote{An input is in general position if in the weighted, directed, and bipartite multigraph with one node per bidder $i$, one node per item $j$, and one node for the dummy item $j_0$ and forward edges from $i$ to $j$ with weight $-v_{i,j}$, backward edges from $j$ to $i$ with weight $v_{i,j}$, reserve-price edges from $i$ to $j$ with weight $v_{i,j}-r_{i,j}$, maximum-price edges from $i$ to $j$ with weight $m_{i,j} - v_{i,j}$, and terminal edges from $i$ to $j_0$ with weight $0$ no two walks that start with the same bidder, alternate between forward and backward edges, and end with a distinct edge that is either a reserve-price edge, a maximum-price edge, or a terminal edge have the same weight.}\textsuperscript{,}\footnote{The example is not in general position because the walk that consists of the maximum-price edge from bidder $1$ to item $1$ and the walk that consists of the forward edge from bidder $1$ to item $1$, the backward edge from item $1$ to bidder $2$, and the maximum-price edge from bidder $2$ to item $1$ have the same weight.}

\begin{example}\label{ex:1}
There are two bidders and one item. The valuations and maximum prices are as follows: $v_{1,1} = 10$, $v_{2,1} = 10$, and $m_{1,1} = m_{2,1} = 5.$ While $\mu=\{(1,1)\}$ with $p_1 = 5$ is ``best'' for bidder 1, $\mu=\{(2,1)\}$ with $p_1 = 5$ is ``best'' for bidder 2. With our definitions a bidder optimal outcome is $\mu = \emptyset$ with $p_1 = 5$. 
\begin{figure}[h]
\centering\scalebox{0.8}{\input{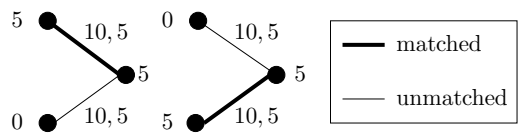}}
\caption{Bidders are on the left side and items are on the right side of the graphs. The numbers next to the bidder indicate his utility, the numbers next to the item indicate its price. The labels along the edge show valuations and maximum prices. Matched edges are bold, while unmatched edges are thin.}
\label{fig:fig1}
\vspace{-10pt}
\end{figure}
\end{example}

The sponsored search market was considered by \cite{ABH09}, who proved the existence of a unique feasible, envy free, and Pareto efficient outcome. They also presented an incentive compatible mechanism to compute such an outcome in polynomial time. Their model, however, is less general than the model studied here as (1) the valuations must be of the form $v_{i,j} = \alpha_j \cdot v_i$, and (2) the maximum prices are per-bidder, i.e., for each bidder $i$ there exists $m_i$ such that $m_{i,j} = m_i$ for all $j$, and are required to be distinct.

Matching markets with more general, non-linear utility functions were studied in \cite{DHW09,CDG10,DHW11}. In \cite{DHW09} we proved the existence of a bidder optimal outcome for general utility functions with multiple discontinuities. In \cite{CDG10} a polynomial-time mechanism for {\em consistent} utility functions with a single discontinuity was given. In \cite{DHW11} we presented a polynomial-time mechanism for piece-wise linear utility functions with multiple discontinuities. 

To summarize: (1) We show how to modify the Hungarian Method in settings with budgets so that it finds a bidder optimal outcome in polynomial time. (2) We show that in settings with budgets no mechanism that computes a bidder optimal outcome can be incentive compatible for all inputs. (3) We show how to extend these results to more general (but still linear) utility functions.

\section{Problem Statement}\label{sec:problem}
We are given a set $I$ of $n$ bidders and a set $J$ of $k$ items. We use letter $i$ to denote a bidder and letter $j$ to denote an item. For each bidder $i$ and item $j$ we are given a {\em valuation} $v_{i,j}$, for each item $j$ we are given a {\em reserve price} $r_{j}$, and for each bidder $i$ and item $j$ we are given a {\em maximum price} $m_{i,j}.$ We assume that the set of items contains a {\em dummy item} $j_0$ for which all bidders have a valuation of zero, a reserve price of zero, and a maximum price of $\infty.$\footnote{Reserve utilities, or {\em outside options} $o_i$, can be modelled by setting $v_{i,j_0} = o_i$ for all $i.$} 

We want to compute an {\em outcome} $(\mu, p)$ consisting of a matching $\mu \subseteq I \times J$ and per-item prices $p=(p_1,\dots,p_k).$ We require that (a) every bidder $i$ appears in exactly one bidder-item pair $(i,j) \in \mu$ and that (b) every non-dummy item $j \neq j_0$ appears in at most one such pair. We allow the dummy item $j_0$ to appear more than once. We call bidders/items that are not matched to any non-dummy item/bidder {\em unmatched}. We regard the dummy item as unmatched, regardless of whether it is matched or not.

The {\em utility} $u_i$ of bidder $i$ is defined as $u_i = 0$ if bidder $i$ is unmatched and it is defined as $u_i = u_{i,j}(p_j)$ if bidder $i$ is matched to item $j$ at price $p_j.$ We set $u_{i,j}(p_j) = v_{i,j} - p_j$ if $p_j < m_{i,j}$ and $u_{i,j}(p_j) = - \infty$ if $p_j \geq m_{i,j}$. We say that the outcome $(\mu,p)$ is {\em feasible} if (1) $u_i \geq 0$ for all $i$, (2) $p_{j_0} = 0$ and $p_j \geq 0$ for all $j \neq j_0$, and (3) $p_j \ge r_j$ for all $(i,j) \in \mu$. We say that a feasible outcome $(\mu,p)$ is {\em envy free} if $u_i \geq u_{i,j}(p_j)$ for all $(i,j) \in I \times J.$\footnote{Since $u_i \ge 0$ and $u_{i,j}(p_j) = - \infty$ if $p_j \ge m_{i,j}$ this is equivalent to requiring $u_i \ge v_{i,j} - p_j$ for all items $j$ with $p_j < m_{i,j}.$} Finally, we say that an envy free outcome $(\mu,p)$ is {\em bidder optimal} if $u_i \geq u'_i$ for all $i$ and envy free outcomes $(\mu',p')$.

We say that a mechanism is {\em incentive compatible} if for every bidder $i$, any two inputs $(v'_{i,j}(\cdot),r_j,m'_{i,j})$ and $(v''_{i,j}(\cdot),r_j,m''_{i,j})$ with (a) $v'_{i,j} = v_{i,j}$ and $m'_{i,j} = m_{i,j}$ for $i$ and all $j$ and (b) $v'_{k,j} = v''_{k,j}$ and $m'_{k,j} = m''_{k,j}$ for $k \neq i$ and all $j$, and corresponding outcomes $(\mu', p')$ and $(\mu'',p'')$ we have that $u_{i,j'}(p'_{j'}) \geq u_{i,j''}(p''_{j''})$ where $(i,j') \in \mu'$ and $(i,j'') \in \mu''$. This formalizes that ``lying does not pay off'' as follows: Even if bidder $i$ claims that his valuation is $v''_{i,j}$ instead of $v_{i,j}$ and that his maximum price is $m''_{i,j}$ instead of $m_{i,j}$ he will not achieve a higher utility with the prices and the matching computed by the mechanism. 

\section{Preliminaries} 

We define the \emph{first choice graph} $G_p = (I \cup J, F_p)$ at prices $p$ as follows: There is one node per bidder $i$, one node per item $j$, and an edge from $i$ to $j$ if and only if item $j$ gives bidder $i$ the highest utility, i.e., $u_{i,j}(p_j) \geq u_{i,j'}(p_{j'})$ for all $j'.$ For $i \in I$ we define $F_p(i) = \{j : \exists \ (i,j) \in F_p\}$ and for $j \in J$ we define $F_p(j) = \{i : \exists \ (i,j) \in F_p\}$. Analogously, for $T \subseteq I$ we define $F_p(T) = \cup_{i \in T} F_p(i)$ and for $S \subseteq J$ we define $F_p(S) = \cup_{j \in S} F_p(j)$. 

\begin{fact}\label{fact:1}
For all prices $p$ such that $p_{j_0} = 0$ and $p_{j} \ge 0$ for all $j \neq j_0$ we have that (1) if $(i,j) \in F_p$ then $p_j < m_{i,j}$, and (2) if the outcome $(\mu,p)$ is envy free then $\mu \subseteq F_p.$
\end{fact}

We define the \emph{feasible first choice graph} $\tilde{G}_p = (I \cup J, \tilde{F}_p)$ at prices $p$ as follows: There is one node per bidder $i$, one node per item $j$, and an edge from $i$ to $j$ if and only if item $j$ gives bidder $i$ the highest utility, i.e., $u_{i,j}(p_j) \geq u_{i,j'}(p_{j'})$ for all $j'$, {\em and} the price of item $j$ is at least the reserve price, i.e., $p_j \geq r_j.$ For $i \in I$ we define $\tilde{F}_p(i) = \{j : \exists \ (i,j) \in \tilde{F}_p\}$ and for $j \in J$ we define $\tilde{F}_p(j) = \{i : \exists \ (i,j) \in \tilde{F}_p\}$. Analogously, for $T \subseteq I$ we define $\tilde{F}_p(T) = \cup_{i \in T} \tilde{F}_p(i)$ and for $S \subseteq J$ we define $\tilde{F}_p(S) = \cup_{j \in S} \tilde{F}_p(i).$ 

\begin{fact}\label{fact:2}
For all prices $p$ such that $p_{j_0} = 0$ and $p_{j} \ge 0$ for all $j \neq j_0$ we have that (1) if $(i,j) \in \tilde{F}_p$ then $r_j \leq p_j < m_{i,j}$, and (2) the outcome $(\mu,p)$ is envy free if and only if $\mu \subseteq \tilde{F}_p.$ 
\end{fact}

We define an {\em alternating path} as a sequence of edges in $\tilde{F}_p$ that alternates between matched and unmatched edges. We require that all but the last item on the path are non-dummy items. The last item can (but does not have to) be the dummy item. A tree in the feasible first choice graph $\tilde{G}_p$ is an {\em alternating tree} rooted at bidder $i$ if all paths from its root to a leaf are alternating paths that either end with the dummy item, an unmatched item, or a bidder whose feasible first choice items are all contained in the tree. We say that an alternating tree with root $i$ is {\em maximal} if it cannot be extended.  

\begin{example}\label{ex:2}
This is a (feasible) first choice graph and a maximal alternating tree for six bidders and six items.
\begin{figure}[h]
\centering\scalebox{0.75}{\input{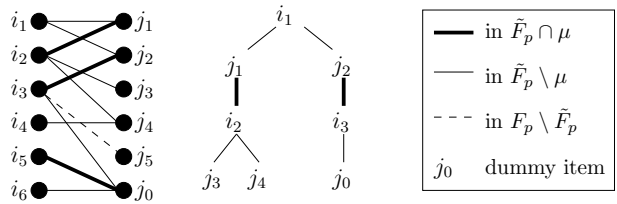}}
\caption{The bidders are $i_1$--$i_6$ and the items are $j_0$--$j_5$. Edges in $\tilde{F}_p \cap \mu$ are thick, edges in $\tilde{F}_p$ are thin, and edges in $F_p \setminus \tilde{F}_p$ are dashed. The (feasible) first choice graph is on the left and the maximal alternating tree is on the right.}
\label{fig:alttree}
\vspace{-10pt}
\end{figure}
\end{example}

\section{Mechanism}\label{sec:hm}

Our mechanism starts with an empty matching and prices all zero. To match an unmatched bidder it computes a maximal alternating tree rooted at this bidder. If at least one of the alternating paths in this tree ends at an unmatched item, then it augments the matching by swapping the matched and unmatched edges along this path. Otherwise it raises the prices of all items that are the first choice of at least one bidder in the tree by $\delta = \min(\delta_{\text{out}},\delta_{\text{res}},\delta_{\text{max}})$, where (a) $\delta_{\text{out}}$ is the minimum amount required to add an edge from a bidder in the tree to an item that was not among his first choice items to the first choice graph, (b) $\delta_{\text{res}}$ is the minimum amount required to turn an infeasible first choice edge that is incident to a bidder in the tree into a feasible first choice edge, and (c) $\delta_{\text{max}}$ is the minimum amount required to make a first choice edge that is incident to a bidder in the tree drop out of the first choice graph.\footnote{For $\delta_{\text{max}}$ to be well defined we need that the utility that bidder $i$ has for item $j$ is defined with $p_j < m_{i,j}$ (weak inequality), and not with $p_j \le m_{i,j}$ (weak inequality).}

While updates corresponding to $\delta_{\text{max}}$ can cause previously matched bidders to become unmatched, neither updates corresponding to $\delta_{\text{out}}$ nor updates corresponding to $\delta_{\text{res}}$ will ever unmatch previously matched bidders: They preserve all feasible first choice edges that are incident to bidders in the tree, and thus the feasible first choice edges along which these bidders are matched. They may cause first choice edges from bidders not in the tree to first choice items of bidders in the tree to drop put of the first choice graph, but none of these bidders will be matched along such an edge (because otherwise the corresponding bidder would belong to the tree).\footnote{This is not true if the reserve prices are allowed to depend on the bidders and the items and the prices are updated as in \cite{DHW10}. See~\ref{app:counterex} for a counter example.}


\begin{verse}
  {\bf Modified Hungarian Method.}\\
  \small
  \vspace{4pt}
  \INPUT valuations $v_{i,j}$, reserve prices $r_j$, maximum prices $m_{i,j}$\\
  \OUTPUT bidder optimal outcome $(\mu,p)$\\
  \vspace{4pt}
  1 \hspace{2pt} $p_j := 0$ for all $j \in J$, $u_i := \max_{j} v_{i,j}$ for all $i \in I$, and $\mu := \emptyset$ \\
  2 \hspace{2pt} \WHILE there exists an unmatched bidder $i_0$ \DO\\
  3 \hspace{2pt} \quad find maximal alternating tree rooted at bidder $i_0$ in $\tilde{G}_p$\\
  4 \hspace{2pt} \quad let $T$ and $S$ be the set of bidders and items in this tree\\
  5 \hspace{2pt} \quad set $u_i := \max_{j} u_{i,j}(p_j)$ for all $i \in T$\\
  6 \hspace{2pt} \quad \WHILE all items $j \in S$ are matched and $j_0 \not \in S$ \DO\\
  7 \hspace{2pt} \quad \quad compute $\delta := \min( \delta_{\text{out}}, \delta_{\text{res}}, \delta_{\text{max}})$ where\\
  8 \hspace{2pt} \quad \quad \quad $\delta_{\text{out}} \ := \min_{i \in T, j \in J \setminus F_p(i)} (u_i + p_j - v_{i,j})$\\
  9 \hspace{2pt} \quad \quad \quad $\delta_{\text{res}} \hspace{4.5pt} := \min_{j \in F_p(T) \setminus \tilde{F}_p(T)} (r_j - p_j)$ \\ 
  10  \quad \quad \quad $\delta_{\text{max}} := \min_{i \in T, j \in F_p(i)} (m_{i,j} - p_j)$\\
  11  \quad \quad update prices, utilities, and matching as follows\\
  12  \quad \quad \quad $p_j := p_j + \delta$ for all $j \in F_p(T)$ \textbackslash \textbackslash \ leads to new graph \\
  13 \quad \quad \quad $u_i := \max_{j} u_{i,j}(p_j)$ for all $i \in T$\\
  14 \quad \quad \quad $\mu \ := \mu \cap \tilde{F}_p$ \textbackslash \textbackslash \ removes unfeasible edges \\
  15 \quad \quad find maximal alternating tree rooted at bidder $i_0$ in $\tilde{G}_p$\\
  16 \quad \quad let $T$ and $S$ be the set of bidders and items in this tree\\
  17 \quad \quad set $u_i := \max_{j} u_{i,j}(p_j)$ for all $i \in T$\\
  18 \quad \ENDWHILE \\
  19 \quad augment $\mu$ along alternating path rooted at $i_0$ in $\tilde{G}_p$\\
  20 \hspace{1pt} \ENDWHILE \\
  21 \hspace{1pt} output $\mu$ and $p$
\end{verse}


\section{Feasibility and Envy Freeness}\label{sec:fs}

\begin{theorem}\label{thm:feasibleandstable}
The Modified Hungarian Method finds a feasible and envy free outcome. It can be implemented to run in time $O(n \cdot k^3)$.
\end{theorem}
\begin{proof}
Since the outcome $(\mu,p)$ maintained by the Modified Hungarian Method satisfies $p_{j_0} = 0$ and $p_{j} \ge 0$ for all $j \neq j_0$ and $\mu \subseteq \tilde{F}_p$ at all times it suffices to show that after $O(n \cdot k^3)$ steps all bidders are matched. The prices, the utilities, and the matching can be initialized in time $O(n \cdot k)$ (l.~1). To analyze the remaining running time we divide it into the total time spent in (1) the outer while loop {\em without} the inner while loop (ll.~2--5 and 19--20) and (2) the inner while loop (ll.~6--18). 

To (1): We have that (a) after each execution of the outer while loop a previously unmatched bidder gets matched and (b) a matched bidder $i$ can only become unmatched if the price of the item $j$ he is matched to reaches $m_{i,j}$. Since there are $O(n)$ many bidders and (b) can happen at most $O(n \cdot k)$ times (a) and (b) show that after $O(n \cdot k)$ executions of the outer while loop all bidders are matched. 
The maximal alternating tree, the utilities of the bidders in the tree, and the augmenting path can be computed via breadth-first search: Start with $T = S = \emptyset$. Add bidder $i_0$ to $T$. Compute the utility of the bidders added to $T$, determine their feasible first choice items, and add these items to $S$. This takes time $O(k)$ per bidder. For each non-dummy item added to $S$ add to $T$ the bidder that this item is matched to. This takes time $O(k)$ per item. Continue like this until no more bidders resp.~items are added to $T$ resp.~$S$. Overall this procedure takes at most $O(k^2)$ steps because at most $O(k)$ bidders resp.~items are added to $T$ resp.~$S$.
We conclude that the total running time of the outer while loop is $O(n \cdot k^3).$

To (2): We say that an iteration of the inner while loop is {\em special} if (a) right before the iteration of the inner while loop the outer while loop was executed, (b) in the previous iteration of the inner while loop a reserve price was reached, or (c) in the previous iteration of the inner while loop a maximum price was reached. Since (a)--(c) can happen at most $O(n \cdot k)$ times the number of special iterations is $O(n \cdot k).$ We show next how a sequence consisting of a special iteration and all non-special iterations that follow it can be implemented in time $O(k^2).$ Since there are at most $O(n \cdot k)$ special iterations this shows that the total running time of the inner while loop is $O(n \cdot k^3).$

The implementation keeps track of $u_i$ for all $i \in T$, $p_j$ for all $j \in F_p(T)$, the matching $\mu$, and the sets $T$ and $S$. In addition, it keeps track of the following slack variables, which have to be initialized at the beginning of the first iteration:
\begin{align*}
  &\gamma^{\text{out}}_{j} = \textstyle{\min_{i \in T}} (u_i + p_j - v_{i,j})
  && \text{for $j \in J \setminus F_p(T)$}\\
  &\gamma^{\text{res}}_{j} = r_j - p_j  
  && \text{for $j \in F_p(T) \setminus \tilde{F}_p(T)$}\\
  &\gamma^{\text{max}}_{j} = \textstyle{\min_{i \in T}} (m_{i,j} - p_j) 
  && \text{for $j \in F_p(T)$}
\end{align*}
Since there are at most $O(k)$ items in $J$ and at most $O(k)$ bidders in $T$ initializing the slack variables takes time $O(k^2)$. Since $\delta_{\text{out}}$, $\delta_{\text{res}}$, and $\delta_{\text{max}}$ are the minima over the corresponding slack variables the slack variables can be used to compute $\delta$ in time $O(k)$. 

We begin by showing how to update the data in all iterations in the sequence with $\delta = \delta_{\text{out}}$. We update the utilities $u_i$ of all bidders $i \in T$ by subtracting $\delta$ in time $O(k).$ We update the prices of all items $j \in F_p(T)$ by adding $\delta$ in time $O(k)$. We do not have to update the matching. 
We update the maximal alternating tree and the utilities of the bidders that are added to the tree as follows: We first add all items that are added to $\tilde{F}_p(T)$ to $S$. Since these are precisely the items for which $\delta = \gamma^{\text{out}}_j$ and $p_j \ge r_j$ we can find these items in time $O(k)$. For each non-dummy item added to $S$ we add to $T$ the bidder that this item is matched to. This takes time $O(k)$ per item. Afterwards we update the utilities of the bidders added to $T$ and add their feasible first choice items to $S$ (in time $O(k)$ per bidder) and for each non-dummy item added to $S$ we add to $T$ the bidder that this item is matched to (in time $O(k)$ per item). We continue like this until no more bidders resp.~items are added to $T$ resp.~$S$. Let $T$ and $S$ resp.~$T'$ and $S'$ denote the sets of bidders and items before resp.~after the update. Then overall this procedure takes at most $O(k + |T' \setminus T| \cdot k + |S' \setminus S| \cdot k)$ steps. 

We update the slack variables as follows. Let $p$ and $p'$ denote the prices before and after the update. For $j \in J \setminus F_p(T)$ and $j \in J \setminus F_{p'}(T')$ we set $\gamma^{\text{out}}_j = \min(\gamma^{\text{out}}_j - \delta, \min_{i \in T' \setminus T}(u_i+p'_j-v_{i,j})).$ This takes time $O(|T'\setminus T|)$ per item. For $j \in J \setminus F_p(T)$ and $j \in F_{p'}(T')$ we remove $\gamma^{\text{out}}_{j}$. If $j \in F_{p'}(T') \setminus \tilde{F}_{p'}(T')$ we add $\gamma^{\text{res}}_j = r_j - p'_j$ and $\gamma^{\text{max}}_j = \min_{i \in T'} (m_{i,j} - p'_j)$. Otherwise, if $j \in \tilde{F}_{p'}(T')$ we only add $\gamma^{\text{max}}_j = \min_{i \in T'} (m_{i,j} - p'_j)$. Removing $\gamma^{\text{out}}_j$ takes time $O(1)$ per item, adding $\gamma^{\text{res}}_j$ takes time $O(1)$ per item, and adding $\gamma^{\text{max}}_j$ takes time $O(k)$ per item. For $j \in F_p(T) \setminus \tilde{F}_p(T)$ and $j \in F_{p'}(T') \setminus \tilde{F}_{p'}(T')$ we update $\gamma^{\text{res}}_j = \gamma^{\text{res}}_j - \delta$ and $\gamma^{\text{max}}_j = \min(\gamma^{\text{max}}_j - \delta, \textstyle{\min_{i \in T' \setminus T}} (m_{i,j} - p'_j))$. For $j \in \tilde{F}_p(T)$ and $j \in \tilde{F}_{p'}(T')$ we update $\gamma^{\text{max}}_j = \min(\gamma^{\text{max}}_j - \delta, \textstyle{\min_{i \in T' \setminus T}} (m_{i,j} - p'_j))$. In both cases, updating $\gamma^{\text{res}}_j$ takes time $O(1)$ per item and updating $\gamma^{\text{max}}_j$ takes time $O(|T' \setminus T|)$ per item.

Since in a sequence of iterations with $\delta = \delta_{\text{out}}$ (a) every iteration adds at least one item to $F_p(T)$, (b) every item can move from $J \setminus F_p(T)$ to $F_p(T)$ at most once, and (c) at most $O(k)$ bidders resp.~items are added to $T$ resp.~$S$ we conclude that updating the data in all iterations in the sequence with $\delta = \delta_{\text{out}}$ takes time $O(k^2).$

We conclude by showing how to update the data in iterations corresponding to $\delta = \delta_{\text{res}}$ or $\delta = \delta_{\text{max}}$ in time $O(k^2)$. The utilities $u_i$ of all bidders $i \in T$ can be updated in time $O(k)$ per bidder by setting $u_{i} = \max_{j} u_{i,j}(p_j)$. The prices $p_j$ of all items $j \in F_p(T)$ can be updated in time $O(1)$ per item by adding $\delta$. The matching $\mu$ can be updated in time $O(k)$ by removing edges $(i,j)$ for which the new price of item $j$ exceeds $m_{i,j}$. 
The maximal alternating tree and the utilities of the bidders in the tree can be computed from scratch in time $O(k^2)$ via breadth-first search (as in the outer while loop).
\end{proof}

\section{Bidder Optimality}\label{sec:bo}

\begin{theorem}\label{the:main}
The Modified Hungarian Method finds a bidder optimal outcome.
\end{theorem}

We proceed as follows: In Lemma~\ref{lem:smallest=optimal} we show that an envy free outcome $(\mu, p)$ is bidder optimal if we have that $p_j \leq p'_j$ for all items $j$ and all envy free outcomes $(\mu', p')$. Afterwards, we define {\em strict overdemand} and prove a lower bound on the price increase of strictly overdemanded items in Lemma~\ref{lem:overdemand}. Finally, in Lemma~\ref{lem:induction}, we argue that whenever the Modified Hungarian Method updates the prices it updates the prices according to Lemma~\ref{lem:overdemand}. This completes the proof.

\begin{lemma}\label{lem:smallest=optimal}
If the outcome $(\mu, p)$ is envy free and $p_j \leq p'_j$ for all $j$ and all envy free outcomes $(\mu', p')$, then the outcome $(\mu, p)$ is bidder optimal.
\end{lemma}
\begin{proof}
For a contradiction suppose that there exists an envy free outcome $(\mu', p')$ such that $u'_i > u_i$ for some bidder $i.$ Let $j$ be the item that bidder $i$ is matched to in $\mu$ and let $j'$ be the item that bidder $i$ is matched to in $\mu'$. Since $p_{j'} \leq p'_{j'}$ and $p'_{j'} < m_{i,j'}$ we have that $u_{i,j'}(p_{j'}) = v_{i,j'} - p_{j'}$. Since the outcome $(\mu, p)$ is envy free we have that $u_i = u_{i,j}(p_{j}) = v_{i,j} - p_{j} \geq u_{i,j'}(p_{j'}) = v_{i,j'} - p_{j'}.$ It follows that $u'_i = v_{i,j'} - p'_{j'} > u_i = v_{i,j} - p_{j} \geq v_{i,j'} - p_{j'}$ and, thus, $p'_{j'} < p_{j'}$. This gives a contradiction.
\end{proof}

We say that a (possibly empty) set $S\subseteq J \setminus \{j_0\}$ is \emph{strictly overdemanded} for prices $p$ with respect to $T\subseteq I$ if (i) $\tilde{F}_p(T) \subseteq S$ and (ii) $\forall \ R \subseteq S$ and $R \neq \emptyset : |\tilde{F}_p(R) \cap T| > |R|$. Using Hall's Theorem~\cite{H35} one can show that an envy free outcome exists for given prices $p$ such that $p_{j_0} = 0$ and $p_{j} \ge 0$ for all $j \neq j_0$ if and only if there is no strictly overdemanded set of items $S$ in the feasible first choice graph $\tilde{G}_p.$ 

\begin{lemma}\label{lem:overdemand}
Given $p$ such that $p_{j_0} = 0$ and $p_{j} \ge 0$ for all $j \neq j_0$ let $u_{i} = \max_{j} u_{i,j}(p_j)$ for all $i.$ Suppose that $S \subseteq J \setminus \{j_0\}$ is strictly overdemanded for prices $p$ with respect to $T \subseteq I$ and let $\delta = \min(\delta_{\text{out}}, \delta_{\text{res}}, \delta_{\text{max}})$, where 
\begin{align*}
&\delta_{\text{out}} \hspace{2pt} = \textstyle{\min_{i \in T, j \in J \setminus F_p(i)}} (u_i + p_j - v_{i,j}), &&\text{and}\\
&\delta_{\text{res}} \ = \textstyle{\min_{i \in T, j \in F_p(i) \setminus \tilde{F}_p(i)}} (r_j - p_j), &&\text{and} \\
&\delta_{\text{max}} = \textstyle{\min_{i \in T, j \in F_p(i)}} (m_{i,j} - p_j).
\end{align*}
Then, for every envy free outcome $(\mu', p')$ with $p'_j \geq p_j$ for all $j$, we have that $p'_j \geq p_j + \delta$ for all $j \in F_p(T).$ 
\end{lemma}
\begin{proof}
We prove the claim in two steps. In the first step, we show that $p'_j \geq p_j + \delta$ for all $j \in \tilde{F}_p(T)$. In the second step, we show that $p'_j \geq p_j + \delta$ for all $j \in F_p(T) \setminus \tilde{F}_p(T)$.

{\em Step 1.} Consider the set of items $A = \{ j \in \tilde{F}_p(T) \ | \ \forall k \in \tilde{F}_p(T): p'_j - p_j \leq p'_{k} - p_{k} \} \subseteq S$ and the set of bidders $B = \tilde{F}_p(A) \cap T \subseteq T.$ If $A = \emptyset$ then there is nothing to show. If $A \neq \emptyset$ then assume by contradiction that $\delta' = \min_{j \in \tilde{F}_p(T)} (p'_j - p_j) < \delta.$ We show below that in this case $|B| > |A|$ and $A \supseteq \tilde{F}_{p'}(B)$. On the one hand this shows that $|A| \ge |\tilde{F}_{p'}(B)|$ and, thus, $|B| > |\tilde{F}_{p'}(B)|.$ On the other hand this shows that $\tilde{F}_{p'}(B) \subseteq A \subseteq S \subseteq J \setminus \{j_0\}$, i.e., $\tilde{F}_{p'}(B)$ does {\em not} contain the dummy item. But if $\tilde{F}_{p'}(B)$ does {\em not} contain the dummy item then the outcome $(\mu', p')$ can only be envy free if every bidder in $B$ is matched to a {\em distinct} item in $\tilde{F}_{p'}(B)$ and, thus, $|B| \le |\tilde{F}_{p'}(B)|$. This gives a contradiction.

The set of items $S$ is strictly overdemanded for prices $p$ with respect to $T$. Thus, since $A \subseteq S$ and $A \neq \emptyset$, we have $|B| = |\tilde{F}_p(A) \cap T| > |A|.$ Next we show that $A \supseteq \tilde{F}_{p'}(B)$. It suffices to show that $\tilde{F}_{p'}(i) \setminus A = \emptyset$ for all bidders $i \in B.$ For a contradiction suppose that there exist a bidder $i \in B$ and an item $k \in \tilde{F}_{p'}(i) \setminus A$. It follows that (1) $u_{i,k}(p'_k) \geq 0$, (2) $u_{i,k}(p'_k) \geq u_{i,k'}(p'_{k'})$ for all $k'$, and (3) $p'_k \geq r_k.$ In particular, $r_k \leq p'_k < m_{i,k}$ and so $u_{i,k}(p'_k) = v_{i,k} - p'_k.$ We also know that there exists $j \in A$ such that $j \in \tilde{F}_p(i)$. Since $j \in A$ we have that $p'_j < p_j + \delta \leq p_j + \delta_{\text{max}} \leq m_{i,j}$ and so $u_{i,j}(p'_j) = v_{i,j} - p'_j$. Thus, since $k \in \tilde{F}_{p'}(i)$, $v_{i,k} - p'_k \geq v_{i,j} - p'_j$. Finally, since $j \in \tilde{F}_p(i)$ and $p_k \leq p'_k < m_{i,k}$, we also have that $u_{i,j}(p_j) = v_{i,j} - p_j \geq u_{i,k}(p_k) = v_{i,k} - p_k$. We distinguish three cases:

{\em Case 1:} $k \in J \setminus F_p(B)$. Since $\delta \leq \delta_{\text{out}} \leq u_i + p_k - v_{i,k}$ and $u_i = v_{i,j} - p_j$ we have that $\delta \leq v_{i,j} - p_j + p_k - v_{i,k}.$ Rearranging this shows that $v_{i,k} - p_k + \delta \leq v_{i,j} - p_j.$ Since $p'_k \geq p_k$ and $p_j > p'_j - \delta$ this implies that $v_{i,k} - p'_k < v_{i,j} - p'_j$. Contradiction!

{\em Case 2:} $k \in F_p(B) \setminus \tilde{F}_p(B)$. We have $\delta \le \delta_{\text{res}} \le r_k - p_k.$ If $p'_k - p_k \leq p'_j - p_j$, then, since $p'_j - p_j = \delta' < \delta$, we have that $p'_k < p_k + \delta \le r_k$. Contradiction! If $p'_k - p_k > p'_j - p_j$, then, since $v_{i,j} - p_j \geq v_{i,k} - p_k$, we get that $v_{i,j} - p'_j > v_{i,k} - p'_k$. Contradiction!

{\em Case 3:} $k \in \tilde{F}_p(B) \setminus A$. Since $j \in A$ and $k \not\in A$ we have that $p'_k - p_k > \delta' = p'_j - p_j.$ Since $v_{i,j} - p_j \geq v_{i,k} - p_k$ this implies that $v_{i,j} - p'_j > v_{i,k} - p'_k$. Contradiction!

{\em Step 2.} Consider an arbitrary item $j \in F_p(T) \setminus \tilde{F}_p(T)$ such that $p'_j - p_j \leq p'_{j'} - p_{j'}$ for all $j' \in F_p(T) \setminus \tilde{F}_p(T)$ and a bidder $i \in T$ such that $j \in F_p(i)$. Assume by contradiction that $\delta' = p'_j - p_j < \delta$. We show that this implies that $\tilde{F}_{p'}(i) = \emptyset$, which gives a contradiction to the fact that the outcome $(\mu', p')$ is envy free.

First observe that $\delta' < \delta \leq \delta_{\text{res}} \leq r_j - p_j$ and, thus, $p'_j < p_j + \delta \leq r_j$, which shows that $j \not\in \tilde{F}_{p'}(i).$ Next consider an arbitrary item $k \neq j.$ For a contradiction suppose that $k \in \tilde{F}_{p'}(i)$. It follows that $r_k \leq p'_k < m_{i,k}$ and $u_{i,k}(p'_k) = v_{i,k} - p'_k \geq u_{i,j}(p'_j)$. Since $p'_j = p_j + \delta' < p_j + \delta \le p_j + \delta_{\text{max}} \leq m_{i,j}$ we have that $u_{i,j}(p'_j) = v_{i,j} - p'_j$ and, thus, $v_{i,k} - p'_k \geq v_{i,j} - p'_j.$ Finally, since $j \in F_p(i)$ and $p_k \leq p'_k < m_{i,k}$, we have that $u_{i,j}(p_j) = v_{i,j} - p_j \geq u_{i,k}(p_k) = v_{i,k} - p_k$.

As in Step 1 we distinguish three cases: If $k \in J \setminus F_p(T)$ or $k \in F_p(T) \setminus \tilde{F}_p(T)$, then by the same argument as in Cases 1 and 2 above we get a contradiction. If $k \in \tilde{F}_p(T)$, then from the result of Step 1 we know that $p'_k - p_k \geq \delta > \delta' = p'_j - p_j$. Since $v_{i,j} - p_j \geq v_{i,k} - p_k$ this implies that $v_{i,j} - p'_j > v_{i,k} - p'_k$, which also gives a contradiction.
\end{proof}

\begin{lemma}\label{lem:induction}
Let $p$ be the prices computed by the Modified Hungarian Method. Then for every envy free outcome $(\mu', p')$ we have that $p_j \leq p'_j$ for all $j.$ 
\end{lemma}
\begin{proof}
We prove the claim by induction over the price updates. Let $p^{t}$ denote the prices after the $t${-th} price update. 

For $t = 0$ the claim follows from the fact that $p_j^{t} = 0$ for all $j$ and $p'_j \geq 0$ for all $j$ and feasible outcomes $(\mu', p')$. 

For $t > 0$ assume that the claim is true for $t-1.$ Let $S \subseteq J \setminus \{j_0\}$ be the set of items and let $T$ be the set of bidders considered by the Modified Hungarian Method for the $t${-th} price update. We claim that $S \subseteq J \setminus \{j_0\}$ is strictly overdemanded for prices $p^{t-1}$ with respect to $T.$ This is true because: (1) $S$ and $T$ are defined as the set of items resp.~bidders in a {\em maximal} alternating tree and, thus, there are no edges in $\tilde{F}_{p^{t-1}}$ from bidders in $T$ to items in $J \setminus S$ which shows that $\tilde{F}_{p^{t-1}}(T) \subseteq S.$ (2) Because for every subset $R \subseteq S$ with $R \neq \emptyset$ all items in $R$ are matched the number of ``neighbors'' that these items have in the maximal alternating tree is strictly larger than $|R|$, i.e., $|\tilde{F}_{p^{t-1}}(R) \cap T| > |R|.$ Since $p^{t-1} \ge 0$ for all $j \in J$ and $p^{t-1}_{j_0} = 0$ and, by the induction hypothesis, $p'_j \geq p^{t-1}_j$ for all $j \in J$ Lemma~\ref{lem:overdemand} shows that $p'_j \geq p^{t-1}_j + \delta$ for all items $j \in F_{p^{t-1}}(T)$. The Modified Hungarian Method sets $p^{t}_j = p^{t-1}_j + \delta$ for all items $j \in F_{p^{t-1}}(T)$ and $p^{t}_j = p^{t-1}_j$ for all items $j \not\in F_{p^{t-1}}(T)$. We conclude that $p'_j \geq p^{t}_j$ for \emph{all} items $j \in J$. 
\end{proof}

\section{Incentive Compatibility}\label{sec:truth}

The following example shows that no mechanism that computes a bidder optimal outcome is incentive compatible for all inputs. In subsequent work we show that every mechanism that computes a bidder optimal outcome is incentive compatible for inputs in general position \cite{DHW11}. Thus, our mechanism---just as the mechanism of \cite{AMPP09}---is incentive compatible for inputs in general position. Note that the example shows that a bidder can improve his utility by lying only about the {\em valuation} of a single item. Also note that (i) there are no reserve prices, i.e., $r_{j} = 0$ for all $j$, (ii) the maximum prices depend only on the item, i.e., for all $i$ there exists a constant $m_i$ such that $m_{i,j} = m_i$ for all $j$, and (iii) no two bidders have the same maximum price, i.e., $m_i \neq m_k$ for any two bidders $i \neq k.$ 

\begin{example}\label{ex:3}
There are three bidders and three items. The valuations are $v_{1,1} = 6$, $v_{1,2} = 5$, $v_{2,1} = 11$, $v_{2,2} = 5$, $v_{2,3} = 4$, $v_{3,2} = 10$, and $v_{3,3} = 4$. The maximum prices are $m_1 = 6$, $m_2 = 4$, and $m_3 = 3$. Reserve prices are zero.
\begin{figure}[h!]
\centering\scalebox{0.8}{\input{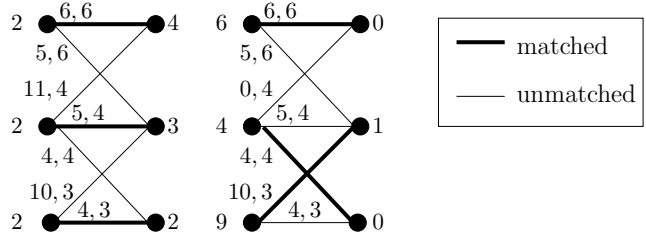}}
\caption{Bidders are on the left side and items are on the right side of the graphs. The numbers next to the bidders indicate their utilities. The numbers next to the items indicate their prices. The labels along the edges show valuations and maximum prices. The graph on the left depicts the bidder optimal outcome for the ``true'' valuations. The graph on the right depicts the bidder optimal outcome for the ``falsified'' valuations. Specifically, in the graph on the right bidder 2 misreports his valuation for item 1. This gives him a strictly higher utility, and shows that lying ``pays off''.}
\vspace{-10pt}
\end{figure}
\end{example}

\section{Generalized Linear Utility Functions}\label{sec:generalization}

The following theorem generalizes our results to utilities of the form $u_{i,j}(p_j) = v_{i,j} - c_i \cdot c_j \cdot p_j$ for $p_j < m_{i,j}$ and $u_{i,j}(p_j) = - \infty$ otherwise. This reduction does {\em not} work if $u_{i,j}(p_j) = v_{i,j} - c_{i,j} \cdot p_j$ for $p_j < m_{i,j}$ and $u_{i,j}(p_j) = - \infty$ otherwise. We give a polynomial-time mechanism for utility functions of this form in~\cite{DHW11}.

\begin{theorem}
The outcome $(\hat{\mu}, \hat{p})$ is bidder optimal for $\hat{v}=(\hat{v}_{i,j})$, $\hat{r} = (\hat{r}_j)$, $\hat{m} = (\hat{m}_{i,j})$ and utility functions $u_{i,j}(p_j) = v_{i,j} - c_i \cdot c_j \cdot p_j$ if $p_j < m_{i,j}$ and $u_{i,j}(p_j) = - \infty$ otherwise if and only if the outcome $(\mu, p)$, where $\mu = \hat{\mu}$ and $p = (c_j \cdot \hat{p}_j)$, is bidder optimal for $v = (\hat{v}_{i,j}/c_i)$, $r = (c_j \cdot \hat{r}_j)$, $m = (c_j \cdot \hat{m}_{i,j})$ and utility functions $u_{i,j}(p_j) = v_{i,j} - p_j$ if $p_j < m_{i,j}$ and $u_{i,j}(p_j) = - \infty$ otherwise. 
\end{theorem}
\begin{proof}
Since $\hat{p}_j < \hat{m}_{i,j}$ if and only if $p < m_{i,j}$ we have that $\hat{u}_{i,j}(\hat{p}_j) = c_i \cdot u_{i,j}(p_j).$ Since $\hat{\mu} = \mu$ this implies that $\hat{u}_i = c_i \cdot u_i$ for all $i.$

{\em Feasibility.}
Since $c_i > 0$ for all $i$ and $c_j > 0$ for all $j$ we have that $\hat{u}_i \geq 0$ for all $i$, $\hat{p}_{j_0} = 0$ and $\hat{p}_j \geq 0$ for all $j$ if and only if $u_i = \hat{u}_i / c_i \geq 0$ for all $i$, $p_{j_0} = c_j \cdot \hat{p}_{j_0}=0$ and $p_j = c_j \cdot \hat{p}_j \geq 0$ for all $j.$ Since $\mu = \hat{\mu}$, $r_j = c_j \cdot \hat{r}_j$, and $p_j = c_j \cdot \hat{p}_j$ for all $i$ and $j$ we have that $\hat{r}_j \leq \hat{p}_j$ for all $(i,j) \in \hat{\mu}$ if and only if $r_j \leq p_j$ for all $(i,j) \in \mu$.

{\em Envy freeness.}
If $(\hat{\mu}, \hat{p})$ is envy free then $(\mu, p)$ is envy free because $u_i = c_i \cdot \hat{u}_i \geq c_i \cdot \hat{u}_{i,j}(\hat{p}_j) = u_{i,j}(p_j)$ for all $i$ and $j.$ If $(\mu, p)$ is envy free then $(\hat{\mu}, \hat{p})$ is envy free because $\hat{u}_i = u_i / c_i \geq u_{i,j}(p_j) / c_i = \hat{u}_{i,j}(\hat{p}_j)$ for all $i$ and $j.$ 

{\em Bidder optimality.}
Suppose that $(\hat{\mu}, \hat{p})$ is bidder optimal but $(\mu, p)$ is not. Then there must be an envy free outcome $(\mu', p')$ such that $u'_i > u_i$ for at least one $i.$ By transforming $(\mu', p')$ into $(\hat{\mu}', \hat{p}')$ we get an envy free outcome for which $\hat{u}'_i = c_i \cdot u'_i > c_i \cdot u_i = \hat{u}_i.$ Contradiction!

Suppose that $(\mu, p)$ is bidder optimal but $(\hat{\mu}, \hat{p})$ is not. Then there must be an envy free outcome $(\hat{\mu}', p')$ such that $\hat{u}'_i > \hat{u}_i$ for at least one $i.$ By transforming $(\hat{\mu}', \hat{p}')$ into $(\mu', p')$ we get an envy free outcome for which $u'_i = \hat{u}'_i / c_i > \hat{u}_i / c_i = u_i.$ Contradiction!
\end{proof}

\section*{Acknowledgements}
We would like to thank Veronika Loitzenbauer and the anonymous referees for their valuable comments.

This work was funded by the Vienna Science and Technology Fund (WWTF) through project ICT10-002 and an EURYI Award.

\bibliographystyle{abbrvnat}
\bibliography{bibliography}

\appendix

\section{Counter Example}\label{app:counterex}
\begin{example}\label{ex:4}
There are three bidders and three items. The valuations are $v_{1,1} = 4$ and $v_{1,2} = v_{2,2} = v_{2,3} = v_{3,2} = v_{3,3} = 6$. The reserve prices are $r_{1,1} = r_{1,2} = r_{2,3} = r_{3,3} = 0$ and $r_{2,2} = r_{3,2} = 4$. All other valuations and reserve prices are zero. Maximum prices are infinity.
\begin{figure}[ht!]
\centering\scalebox{0.8}{\input{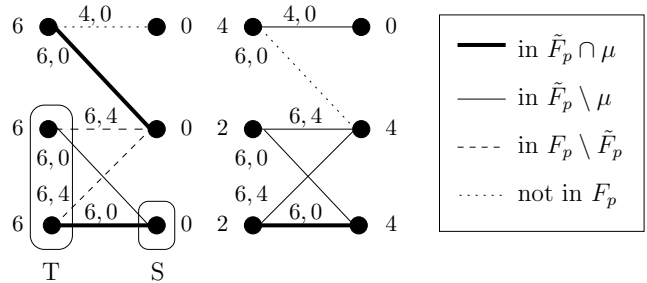}}
\caption{Bidders are on the left side and items are on the right side of the graphs. The numbers next to the bidders indicate the utilities that they would get if they were matched to one of their first choice items. The numbers next to the items indicate their prices. The labels along the edges show valuations and reserve prices. Edges in $\tilde{F}_p \cap \mu$ are thick, edges in $\tilde{F} \setminus \mu$ are thin, edges in $F_p \setminus \tilde{F}_p$ are dashed, and all other edges are dotted. 
If the prices of the items in $F_p(T)$ are updated as in \cite{DHW10}, i.e., $\delta_{\text{res}} = \min_{i \in T, j \in F_p(i)\setminus\tilde{F}_p(i)} (r_{i,j}-p_j)$, then bidder $1$ gets unmatched.
This shows that with bidder-item dependent reserve prices $r_{i,j}$ bidders can also get unmatched if no maximum price is reached.}
\vspace{-10pt}
\end{figure}
\end{example}

\end{document}